\documentclass[12pt,review]{elsarticle}

\usepackage{url}
\usepackage{amsthm}

\usepackage[ruled]{algorithm2e}
\usepackage{subfigure}
\usepackage{multirow}
\usepackage{amsmath}
\usepackage{graphicx}
\usepackage{color}
\usepackage{caption}
\usepackage{multicol}
\usepackage{amssymb}
\usepackage{ulem}

\usepackage{booktabs} 
\usepackage{multirow}
\usepackage{xcolor}
\usepackage{subfigure}
\usepackage{tablefootnote}
\usepackage{hyperref}
\usepackage{bm}

\journal{Neurocomputing}

\begin{document}

\begin{frontmatter}



\title{Automatic Self-supervised Learning for Social Recommendations}


\author[SAI]{Xin~He(hexin20@mails.jlu.edu.cn)}
\author[PLU]{Wenqi~Fan(wenqi.fan@polyu.edu.hk)}
\author[CS]{Ying~Wang(wangying2010@jlu.edu.cn)}
\author[CS]{Mingchen~Sun(mcsun20@mails.jlu.edu.cn)}
\affiliation[SAI]{organization={ The Department
of Computer Science and Technology, the School of Artificial Intelligence (SAI), at Jilin University},
            addressline={2699 Qianjin Street},
            city={Changchun},
            state={Jilin},
            country={China}}

\affiliation[PLU]{organization={The Hong Kong Polytechnic University},
            addressline={Kowloon},
            city={Hong Kong},
            state={Hong Kong},
            country={China}}


\affiliation[CS]{organization={The Department of Computer Science and Technology,
College of Computer Science and Technology, at Jilin University},
            addressline={2699 Qianjin Street},
            city={Changchun},
            state={Jilin},
            country={China}}

\author[SAI]{Xin Wang \corref{mycorrespondingauthor}}
\cortext[mycorrespondingauthor]{Corresponding author}
\ead{xinwang@jlu.edu.cn}

\begin{abstract}
In recent years, researchers have leveraged social relations to enhance recommendation performance. 
However, most existing social recommendation methods require carefully designed auxiliary social tasks tailored to specific scenarios, which depend heavily on domain knowledge and expertise. 
To address this limitation, we propose Automatic Self-supervised Learning for Social Recommendations (AusRec), which integrates multiple self-supervised auxiliary tasks with an automatic weighting mechanism to adaptively balance their contributions through a meta-learning optimization framework. 
This design enables the model to automatically learn the optimal importance of each auxiliary task, thereby enhancing representation learning in social recommendations.
Extensive experiments on several real-world datasets demonstrate that AusRec consistently outperforms state-of-the-art baselines by 3.3\%–10.7\% in Recall@10 and 1.4\%–7.1\% in NDCG@10, validating its effectiveness and robustness across different recommendation scenarios. 
The code is in: \href{https://github.com/hexin5515/AusRec}{https://github.com/hexin5515/AusRec}.
\end{abstract}



\end{frontmatter}


\section{Introduction}
In the age of information overload, people are increasingly overwhelmed by the vast amount of information available online, which motivates intelligent information systems to help select the relevant information that can match users' personal interests from the huge amount of data. 
As representative intelligent information systems, recommendation systems aim to mitigate information overload and are widely applied across various online platforms~\cite{liu2024privacy,fan2020graph,meng2024poi}, such as online shopping (e.g., Amazon\footnote{\url{https://www.amazon.cn}} and Taobao\footnote{\url{https://www.taobao.com}}) and social media (e.g.,  Facebook\footnote{\url{http://www.facebook.com}}, Twitter\footnote{\url{http://twitter.com}}, and Weibo\footnote{\url{https://weibo.com}}). 
More specifically, recommendation systems perform to suggest a list of items to users that match users' interests and are more likely to be interacted by users. 
As one of the most widely used technologies in existing advanced recommendation systems, collaborative filtering (CF) is based on the assumption that users who have similar preferences in the history are likely to have similar preferences in the future~\cite{mcpherson2001birds}. 
In general, user-item historical interactions are key for expressing users' preferences and can be encoded to learn the representations of both users and items in recommendation systems~\cite{koren2009matrix,fan2019deep_dscf,ni2023enhancing}. 

\begin{figure}[t]
  \centering
  \includegraphics[width=1.0\linewidth]{ 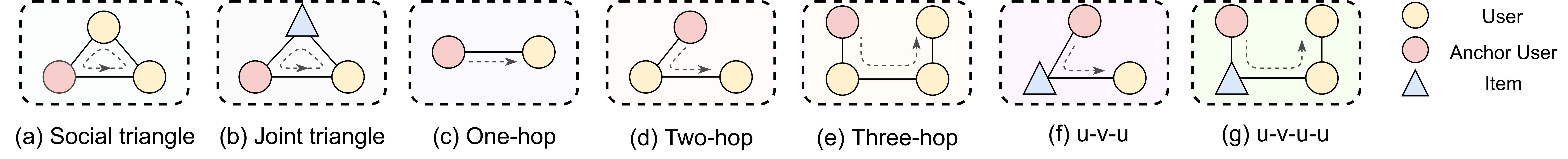}
  \caption{Examples of various social relations in social networks. 
  $u$ indicates a user, and $v$ indicates an item. 
}
\label{high_order}
\end{figure}

Over the past decade, the development of online social networking services (e.g., Facebook,  Weibo, Douban) has been significant, recent studies show that social relations between users can be incorporated into enhancing the representation learning of users
for recommendation systems~\cite{de2024exploiting,fan2020graph,fan2019graph,wu2019neural,yu2020enhance}, which is motivated by the observation that users' preferences are likely to be influenced by their socially connected friends (e.g., classmates, colleagues, family members, etc.)~\cite{kim2007impact}. 
For example, SocialMF~\cite{jamali2010matrix} and GraphRec~\cite{fan2019graph} incorporate first-order social neighbors to enhance user representation learning. DiffNet~\cite{wu2019neural} proposes an influence diffusion method to stimulate social influence propagation by considering high-order social information. ESRF~\cite{yu2020enhance} proposes to utilize triangle relations among users in social recommendations. 
We illustrate some representative social relations tasks in most existing social recommendations in Figure~\ref{high_order}, such as one-hop relations among users,  higher-order relations among users, triangular social relations,  meta-path-based social relations, etc. 
Intuitively, in social networks, users can share more similar preferences towards items among social triangle relations (with strong ties) than meta-path-based social relations (with weak ties).
In summary, most existing social recommendation methods are required to carefully develop various auxiliary social relation tasks (e.g., high-order, social triangle, etc.) to enhance user representation learning by considering the characteristics of social networks/graphs for improving social recommendations~\cite{tang2013social}. 

In recent years, Self-Supervised Learning (SSL) techniques have become an emerging learning paradigm that exploits rich unlabeled data in enhancing representation learning via designing appropriate auxiliary tasks in various research fields, such as CV and NLP~\cite{wu2024self,hashmi2025self}. 
Owing to their powerful capabilities in representation learning with unlabeled data,  self-supervised learning has been successfully incorporated into recommendation systems~\cite{wu2021self,xia2021self,yu2021self,tian2024self}. 
More specifically, most existing SSL-based recommendation methods propose to design a few SSL auxiliary tasks to assist the primary (target) task (i.e., item recommendations).
For example, SGL~\cite{wu2021self} presents a general framework for self-supervised graph learning with stochastic augmentations on user-item graphs (e.g., node/edge dropout) for recommendations.
DHCN~\cite{xia2021self} integrates self-supervised learning into network training by maximizing the mutual information from two channels. 
MHCN~\cite{yu2021self} uses social triangle relations to conduct contrastive learning auxiliary tasks for improving social recommendation.

Despite the aforementioned successes, they do not fully take advantage of self-supervision signals to enhance representation learning under the self-supervised learning paradigm in social recommendations due to their fundamental challenges.
\textbf{First}, the adopted Self-Supervised Auxiliary tasks (SS-A tasks for short) must be carefully selected, requiring substantial domain knowledge and expertise in primary task (e.g., complicated recommendation scenarios and datasets' characteristics), since different SS-A tasks vary in importance depending on the specific datasets and downstream tasks.
\textbf{Second}, inappropriate combinations with multiple SS-A tasks can result in \textbf{negative transfer}, in which various SS-A tasks may dominate training and hinder representation learning in the primary recommendation task. Worse still, balancing multiple SS-A tasks to assist the primary task is immensely challenging. 
However, most existing self-supervised enhanced recommendation methods only incorporate a few SS-A tasks, assign equal weights to various SS-A tasks, or heavily tune hyper-parameters to control the contribution of different SS-A tasks~\cite{wu2021self,xia2021self,yu2021self}. 

\begin{figure*}[t]
\centering
{\subfigure[LastFM]
{\includegraphics[width=0.327\linewidth]{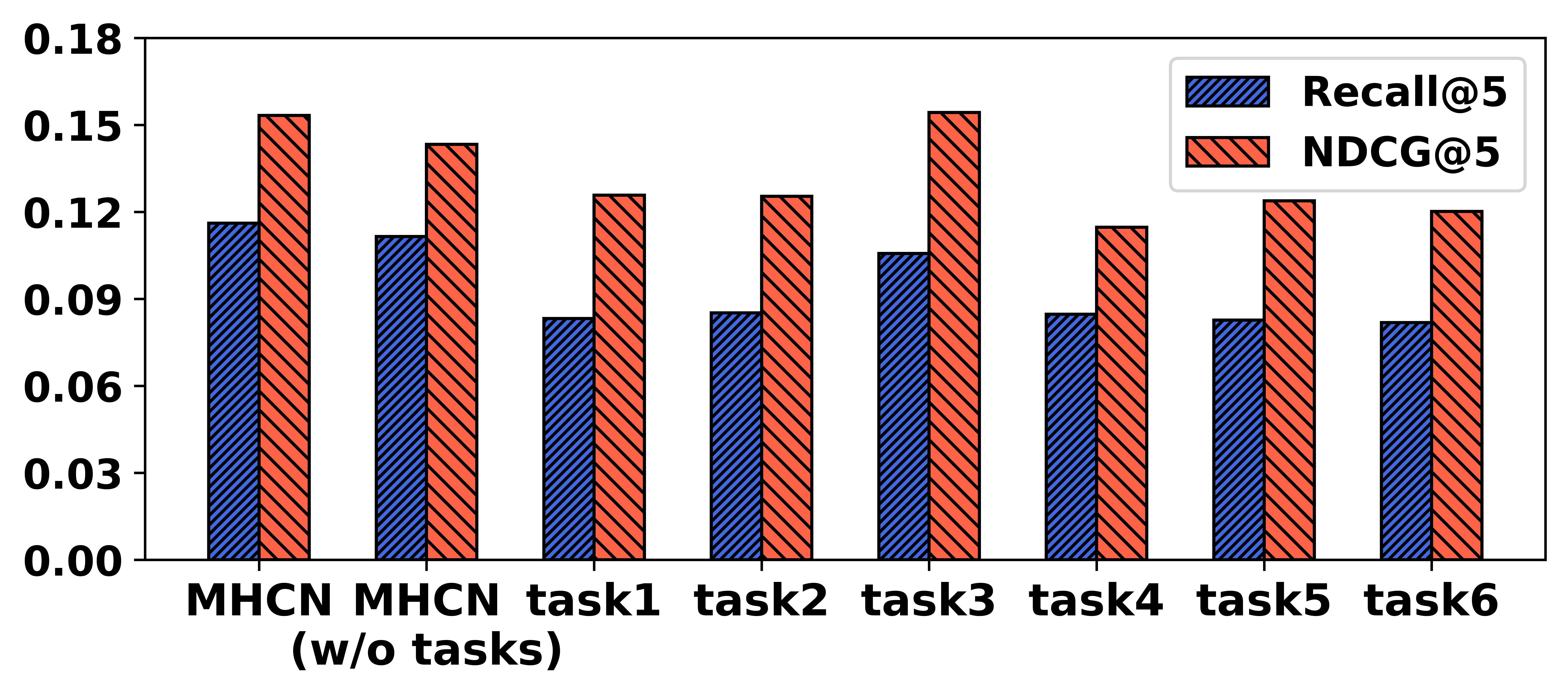}}}
{\subfigure[Epinions]
{\includegraphics[width=0.327\linewidth]{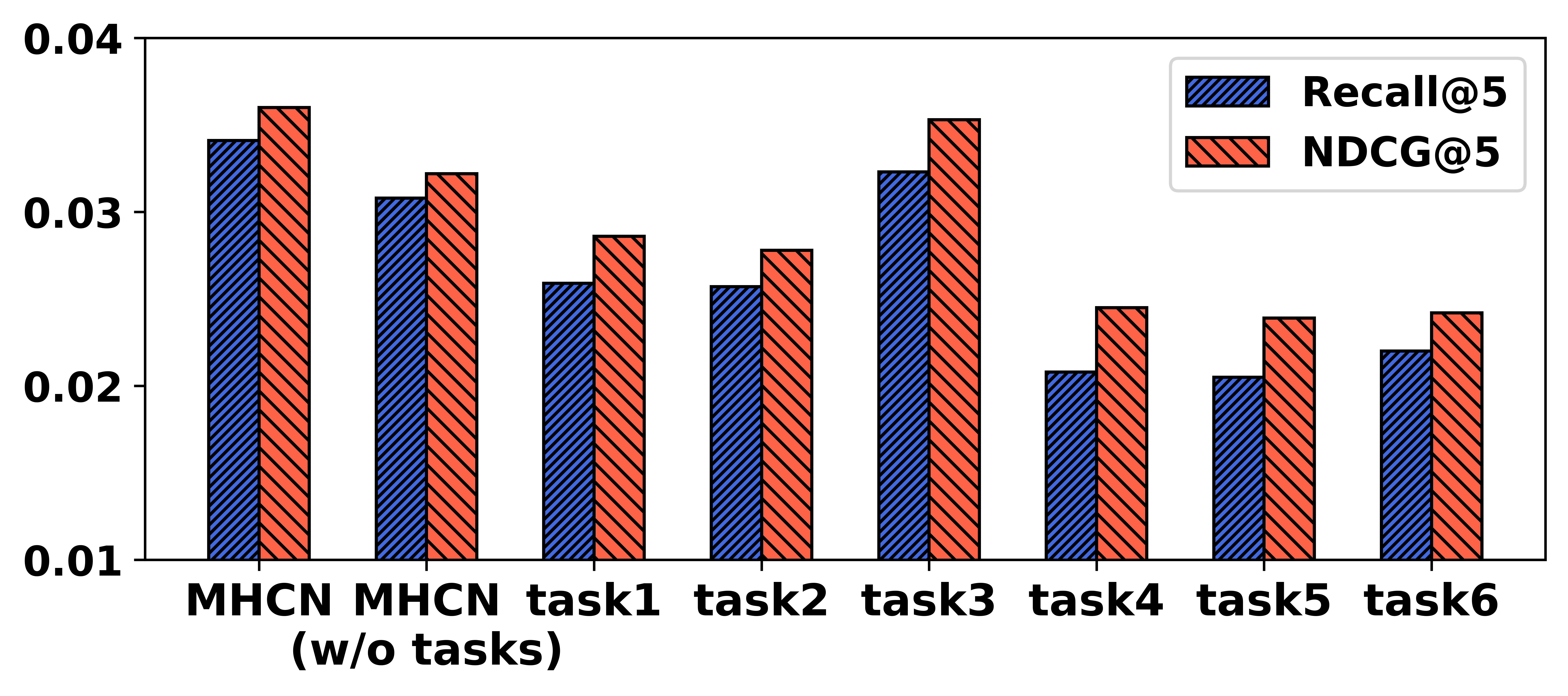}}}
{\subfigure[DBook]
{\includegraphics[width=0.327\linewidth]{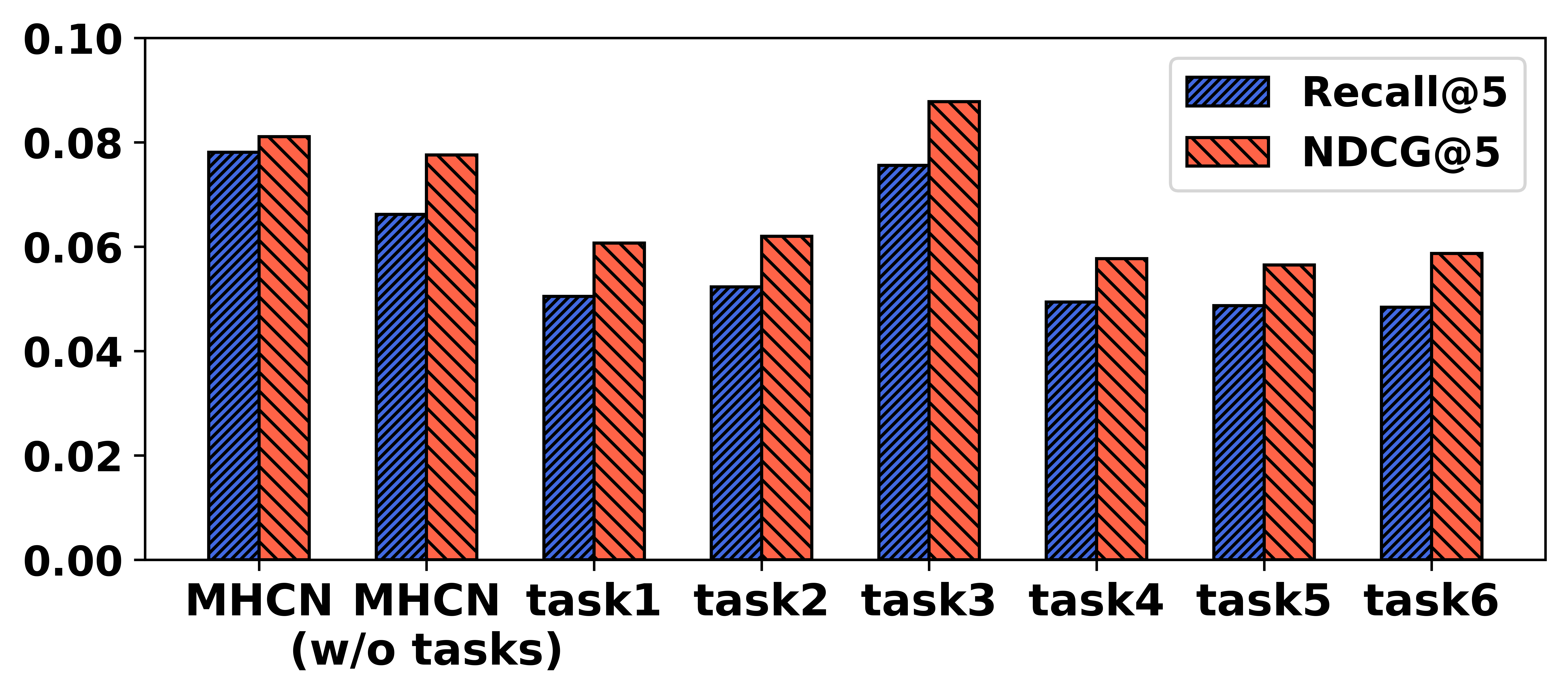}}}%

\caption{Performance of multiple self-supervised auxiliary tasks for the advanced social recommendation method  MHCN~\protect\cite{yu2021self} under Recall@5 and NDCG@5 metrics over three datasets (i.e., LastFM, Epinions, and DBook). MHCN(w/o tasks) denotes the MHCN model without any auxiliary tasks enhanced. 
{The versions of MHCN model with single self-supervised auxiliary task enhanced are represented by ssl 1 to 6, }
including social triangle relation, joint triangle relation, meta-path relation ($u_i \to v_j \to u_{i'}$), 1-hop neighbors, 2-hop neighbors, 3-hop neighbors. 
}
\label{fig:MHCN}
\end{figure*}

In order to verify the aforementioned weaknesses, we conduct preliminary studies to examine how a variety of SS-A tasks perform on an advanced SSL-based social recommendation method MHCN  over three datasets, including LastFM, Epinions, and DBook (refer to Section~\ref{sec:datasets} for more details.).
The experimental results are presented in Figure~\ref{fig:MHCN}, where it can be observed that different SS-A tasks have a distinct influence on the primary recommendation task across different datasets. 
What's worse, MHCN with various SSL tasks (except task 3) imposes a negative effect on MHCN (w/o tasks), where these methods fail to transfer knowledge from multiple SS-A tasks to the primary recommendation task. From these observations, we can conclude that the construction of SS-A tasks in social recommendation strongly depends on the datasets and domain knowledge. 
Therefore, an important research question is \textbf{how to automatically take advantage of multiple SS-A tasks to enhance representation learning for improving social recommendations.}
To the best of our knowledge, the studies to automatically leverage various SS-A tasks for social recommendations remain rarely explored. 

In this paper, to tackle the aforementioned challenges, we aim to build an \textbf{Au}tomatic self-supervised learning for \textbf{S}ocial \textbf{Rec}ommendations (\textbf{AusRec}), which can automatically leverage SS-A tasks for assisting the primary recommendation task. More specifically, we propose an automatic weighting mechanism to learn weights for various SS-A tasks, so as to balance the contribution of such SS-A tasks for enhancing representation learning in social recommendations. What's more, the proposed automatic weighting mechanism is formulated as a meta learning optimization problem with an automatic weighting network. The main contributions are summarized as follows:

\begin{itemize}

\item We study a novel problem of taking advantage of multiple self-supervised social auxiliary tasks by adjusting their weights in an end-to-end manner for social recommendations;

\item We introduce a principled way  to incorporate multiple SS-A tasks, so as to transfer informative knowledge to enhance representation learning by utilizing abundant unlabeled data in social relations;

\item We propose a novel self-supervised learning framework for social recommendations (\textbf{AusRec}), which automatically balance the importance of various SS-A tasks to improve the primary social recommendations;

\item We conduct comprehensive experiments and ablation studies on various real-world datasets to demonstrate the superiority of AusRec.
\end{itemize}

The remainder of this paper is organized as follows. Section 2 reviews related work on recommendation systems, social recommendations, self-supervi-sed learning, and meta-learning for recommendation systems. Section 3 presents the proposed AusRec in detail, including the formulation of self-supervised auxiliary tasks, the automatic weighting mechanism, and the model training process. Section 4 reports the experimental setup, datasets, baseline comparisons, and ablation studies. Section 5 discusses the limitations and feature work of our approach, and Section 6 concludes the paper.


\section{Related Work}
\label{sec:relatedwork}

In this section, we review related work of recommendation systems, social recommendation methods, self-supervised learning, and meta-learning approaches for recommendation systems.

\subsection{Recommendation Systems}
Modern recommendation systems generally fall into three categories: cont-ent-based, collaborative filtering, and hybrid systems. Collaborative filtering (CF) is one of the most widely adopted techniques in these systems, known for its strong performance across various recommendation scenarios.
For example, MF~\cite{koren2009matrix} represents each user and item ID as a vectorized representation, and calculates user's preferences for items by conducting inner product between user and item.
In practice, instead of expressing preferences directly in terms of ratings, users also express their preferences for items implicitly by clicking or browsing. 
Later on, Bayesian Personalized Ranking (BPR)~\cite{rendle2012bpr} is developed to handle implicit feedback between users and items, optimizing a pairwise ranking loss to ensure that items liked by a user rank higher than items disliked by the user in their recommendation list.
In recent years, with the success of representation learning, graph neural networks (GNNs)~\cite{juan2025dynamic,juan2024multi,wang2024two} have been increasingly adopted in recommendation systems~\cite{fan2021graph}. 
LightGCN~\cite{he2020lightgcn} simplifies the GNNs-based recommendation model by removing feature transformation and nonlinear activation, enhancing training efficiency and achieving better performance. 
Thus, it is often considered as a backbone method for new GNNs-based recommendation methods or a baseline method for comparison with new GNNs-based recommendation methods. 

\subsection{Social Recommendations}
With the rise of social media, leveraging social relationships for recommendations has gained significant attention in recent years~\cite{jamali2010matrix,wu2019neural,fan2019graph,yu2020enhance,fan2018deep}. 
In particular, Mohsen Jamali et al. propose a novel matrix factorization-based recommendation method SocialMF~\cite{jamali2010matrix}, which incorporates social influence into the model so that each user's features depend on the feature vectors of their direct neighbours in social networks. As cold-start users are more dependent on social networks than the general users, the impact of using trust propagation on cold start users becomes even more important.
What's more, recent deep learning techniques have advanced classic social recommendation methods~\cite{zhou2024multi,yu2020enhance}. 
More specifically, due to the great success in learning meaningful representations, graph neural networks techniques have significantly improved the performance of social recommendations~\cite{fan2019graph,wu2019neural,yu2020enhance}. 
For example, Fan et al. design the very first method GraphRec\cite{fan2019graph} to employ the power of graph neural networks for improving the learning of user and item representations in social recommendations. 
Wu et al. propose to model the process of social influence propagation via  GNNs model to obtain improved representations of users and items~\cite{wu2019neural}. 
Yu et al. develop a GCN-based auto-encoder to enhance relational data by encoding more intricate higher-order connection patterns~\cite{yu2020enhance}. 
Note that most of the existing social recommendation methods introduce one or a few types of social relations into the recommendation models. 
In this work, we explore incorporating multiple social relations into recommendations and automatically leveraging them for the primary recommendation task.

\subsection{Self-supervised Learning}
In recent year, Self-supervised learning (SSL) is proposed to enhance representation learning by training with ground-truth samples derived from the raw data.
Recent advances in SSL have significantly broadened its applications across various domains, including computer vision~\cite{chen2024deconstructing,assran2023self}, natural language processing~\cite{baevski2022data2vec,chung2021w2v}, and graph representation learning~\cite{peng2020graph,qiu2020gcc,sun2020multi,velickovic2019deep,sankar2020beyond,hwang2020self,wu2021self_survey,jin2020self}.
These studies mainly design SS-A tasks in the view of the data structure
to mark the unlabeled data~\cite{jin2020self}. For instance, 
TimesURL~\cite{chen2024deconstructing} enhances model performance by introducing a frequency-temporal-based augmentation strategy and designing double universums as high-quality hard negatives to improve contrastive learning (A type of SSL).
data2vec~\cite{baevski2022data2vec} improves model performance by introducing a unified self-supervised learning framework that predicts contextualized latent representations of the full input from a masked view across modalities.
InfoMotif~\cite{sankar2020beyond} utilizes correlations among attributes in motifs to regularize graph neural networks with mutual information maximization.


As an emerging paradigm, SSL achieves great success in recommendation systems~\cite{yu2022self,wu2021self,yu2021self,liu2022self}.
SSL in recommendation systems can be divided into four categories: contrastive-based, predictive-based, generative-based, and hybrid~\cite{jaiswal2020survey,wu2021self_survey}. Contrastive Learning (CL)-based methods treat a single user or item as an instance, then generate multiple views based on the same instance via different data augmentation methods, and finally bring views of the same instance closer together while distancing views of different instances in the embedding space.
For example, SGL~\cite{wu2021self} adopts various data augmentation methods to construct SS-A tasks for improving the performance of recommendation systems. DHCN~\cite{xia2021self} integrates SSL into the model training by optimizing mutual information among the representations of sessions. MHCN~\cite{yu2021self} employs social triangle relations to construct contrastive SS-A tasks to improve social recommendations. The generative methods design a SS-A task to reconstruct the original user/item information, making the model learn to predict part of the data based on another part of the data, which is inspired by the masked language models~\cite{devlin2018bert,joshi2020spanbert,conneau2019cross,song2019mass}. Different from generative methods, predictive methods~\cite{yu2021socially} can be seen as a self-prediction rather than predicting the missing parts of raw data. 
The original data is used to produce new samples and labels.
The hybrid methods integrate the SS-A tasks constructed based on the three different methods mentioned above into a recommendation model~\cite{wu2021self_survey}. These methods achieve a more comprehensive self-supervision by integrating different types of self-supervised signals, and further improve the performance of recommendation systems.
Despite the compelling successes, most existing SSL-based recommendation methods only incorporate a few SS-A tasks, assign equal weights to various SS-A tasks, or heavily tune hyper-parameters to regulate the contribution of various SS-A tasks. Meanwhile, they are unable to automatically balance multiple SS-A tasks for assisting target social recommendations across datasets. In this work,  we make the first attempt to fill this gap by introducing an automatic self-supervised learning framework for social  recommendations.

Beyond recommendation systems, self-supervised learning has been widely explored in other domains, such as feature-coupled networks~\cite{wang2025non} and human activity recognition~\cite{yuan2024self}. These approaches demonstrate that carefully designed auxiliary tasks can capture complementary structural and semantic information, leading to more robust and generalizable representations. Inspired by these advances, AusRec automatically integrates multiple self-supervised tasks in a task-aware manner, avoiding manual selection while preserving training and inference efficiency.

\subsection{Meta Learning for Recommendation Systems}
Meta learning has achieved remarkable successes on a variety of deep learning tasks in recent years~\cite{li2024few,li2024contrastive,pan2022multimodal}. 
For example, meta learning plays an important role in few-shot learning, learning a base model that generates the parameters of other models or learning optimizers that can generalize to the new tasks. Besides, meta learning also aims to improve the performance of existing algorithms via metadata~\cite{chen2021autodebias,finn2017model}.
Recently, there are some studies~\cite{lee2019melu,bharadhwaj2019meta,chen2019lambdaopt,zhang2020retrain} have investigated meta learning techniques 
to recommendation systems. MetaCS-DNN~\cite{lee2019melu} is trained in a meta learning manner, so as to achieve the needs of a wide range of users, and the model can also be quickly adapted to the specific user after a few update steps.
MeLU~\cite{bharadhwaj2019meta} attempts to solve the user cold-start problem in recommendation systems from the view of few-shot via meta learning.
$\lambda \textbf{O}_{\textbf{PT}}$~\cite{chen2019lambdaopt} learns finer-grained regularization parameters for recommendation via meta learning.
SML~\cite{zhang2020retrain} improves the performance of recommendation systems by leveraging meta learning to guide the process of re-training.
In our paper, we adopt meta learning techniques to learn automatic weights towards various self-supervised tasks and transfer knowledge in assisting the primary recommendation task.

\section{The Proposed Method}

\label{sec:methodlogy}

This section presents the definitions and notations used throughout the paper, followed by an introduction to our proposed automatic self-supervised learning framework for social recommendations (AusRec). Finally, we detail the model training process and parameter learning approach.

\subsection{Definitions and Notations}

Let $U = \{u_1, u_2, ..., u_m\}$ and $V = \{v_1, v_2, ..., v_n\}$ be the sets of users and items, respectively, where $m$ is the number of users and $n$ is the number of items.
We denote $\mathbf{R}\in\mathbb{R}^{m\times{n}}$ as a user-item interaction  matrix, where $r_{ij}$ = 1 if user $u_i$ has interacted with item $v_j$, otherwise $r_{ij}$ = 0. We use $\mathbf{S}\in\mathbb{R}^{m\times{m}}$ to denote the user-user social  matrix, where $s_{ii'} = 1$ if user $u_i$ has a social relation with user $u_{i'}$, otherwise $s_{ii'} = 0$. 
In this work, since data in social recommendations can be naturally represented as graph structure $\mathcal{G} = (\mathcal{G}^{\mathbf{R}}, \mathcal{G}^\mathbf{S})$, 
the user-item interaction matrix ($\mathbf{R}$) and user-user social network ($\mathbf{S}$) can be  represented as $\mathcal{G}^{\mathbf{R}}$ and $\mathcal{G}^\mathbf{S}$, respectively.
Mathematically, the adjacency matrix of the graph $\mathcal{G}$ which combines user-item graph $\mathcal{G}^{\mathbf{R}}$ and user-user social graph $\mathcal{G}^\mathbf{S}$ can be defined as:

\begin{align}
\label{eq:adj}
    \mathbf{A}\in\mathbb{R}^{(m+n)\times{(m+n)}} = \begin{pmatrix}\mathbf{S}& \mathbf{R}\\\mathbf{R}^{T} & \mathbf{0}\\\end{pmatrix}.
\end{align}

Given $\mathbf{R}$ and $\mathbf{S}$, our goal is to predict users’ unknown preferences towards items (i.e., the unobserved entries in $\mathbf{R}$) by learning user and item representations.

\subsection{An Overview of the Proposed Method}
\begin{figure*}[!t]
  \centering
  \includegraphics[width=1.0\textwidth]{ 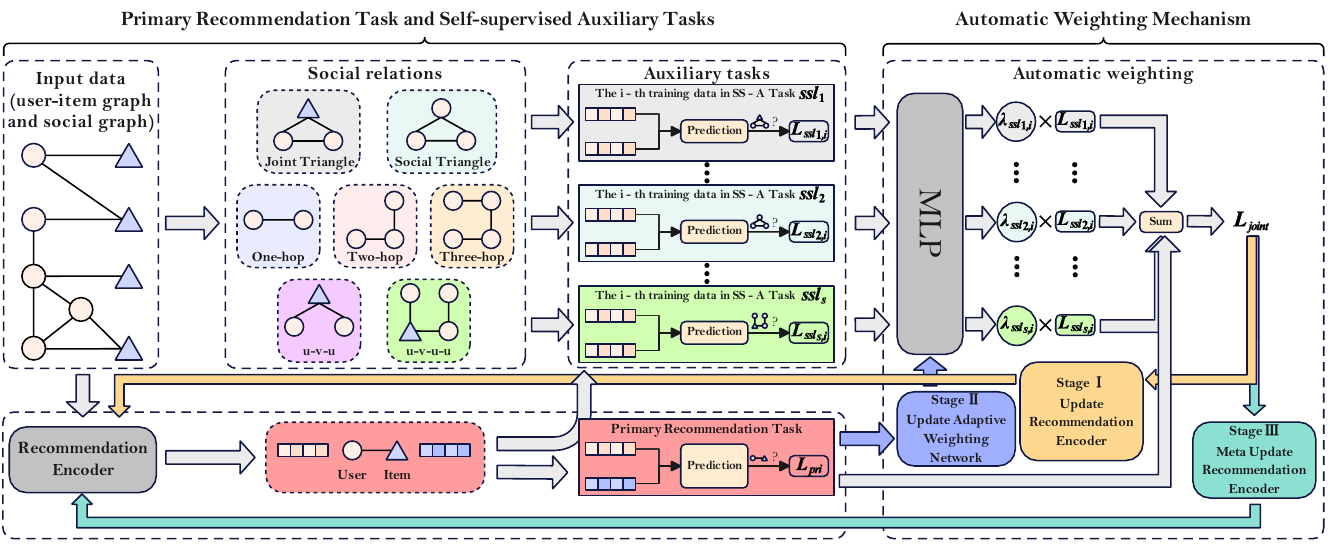}
  \caption{The overall framework of the proposed method.  The parameters updating process consists of three main stages in the automatic weighting mechanism. Stage I performs a hypothetical update of the recommendation encoder ($\hat{\mathbf{W}}$) using the current task weights. Stage II evaluates these weights on a meta dataset and updates the weighting network ($\mathbf{\Theta}$) accordingly. Stage III then updates the encoder ($\mathbf{W}$) with the newly optimized weights.}
\label{fig.2}
\end{figure*}

Figure~\ref{fig.2} illustrates the architecture of the proposed model, comprising two primary components: (1) \textbf{primary recommendation task and self-supervised auxiliary tasks}, which aims to construct multiple SS-A tasks to assist the primary social recommendations; (2) \textbf{automatic weighting mechanism}, which is introduced to balance multiple SS-A tasks by learning the weights towards different SS-A tasks via meta learning.
More specifically, the SS-A tasks assist the primary recommendation task by predicting whether there is a specific social relation between users, aiming to transfer knowledge from the SS-A tasks to the primary recommendation task. 
At the same time, an automatic weighting network is developed to automatically adjust the weights of different SS-A tasks via meta learning. 
The  main goal of our proposed framework can be summarized as:
\begin{itemize}
\item  Learning the representations of users and items from user-item interactions and social networks to make recommendations by constructing various SS-A tasks.
\item  Automatically weighting various SS-A tasks, balancing primary and SS-A tasks to avoid negative knowledge transfer and further improve the performance of the primary recommendation task. 
\end{itemize}
Next, we will detail each model component.

\subsection{Primary Recommendation Task and Self-supervised Auxiliary Tasks}
This component aims to construct multiple SS-A tasks to assist the primary social recommendation task.  First, we introduce the primary recommendation task to model social recommendations. Second, we detail how to take advantage of multiple SS-A tasks for enhancing user and item representation learning. 
Finally, we provide the overall optimization objective of our proposed framework. 

\subsubsection{Primary Recommendation Task: Social Recommendations} 
Due to the success of graph neural networks (GNNs)~\cite{miao2024rethinking} in representation learning, we adopt the GNNs-based recommendations to encode user-item interactions and social relations for user and item representation learning~\cite{he2020lightgcn,wu2019neural}. Here, LightGCN~\cite{he2020lightgcn}, as one of the most representative methods, is adopted as our target encoder to learn user and item representation in user-item graph $ \mathcal{G}^{\mathbf{R}}$ and user-user social graph $\mathcal{G}^\mathbf{S}$ for social recommendations.  
More specifically, given the adjacency matrix $\mathbf{A}\in\mathbb{R}^{(m+n)\times{(m+n)}}$ (in Eq.~\ref{eq:adj}) for modeling graph data $\mathcal{G} = (\mathcal{G}^{\mathbf{R}}, \mathcal{G}^\mathbf{S})$ in social recommendations, node $u$'s representation (i.e., user or item) can be learned by aggregation operation as follows:
\begin{equation}
    \mathbf{e}_u^{k+1} = \text{AGG}(\{\mathbf{e}_v^{k}:v\in\mathcal{N}_u\}),
\end{equation}where $\mathbf{e}_u^{k+1}$ denotes the representation of the user node after aggregating the representations $\mathbf{e}_v^{k}$ of the item nodes connected to it at the $k$-th layer. $\text{AGG}$ denotes the GNNs aggregation function to learn node $u$'s representation through the local neighbors $\mathcal{N}_u$ in $\mathcal{G}$. To be more specific, if ${u_*}$ represents a user node, $\mathcal{N}_{u_i}$ contains the set of items (with which $u_i$ has interacted with in the user-item interaction network $\mathbf{R}$) and the set of social neighbors (whom $u_i$ is directly connected with in the user-user social network $\mathbf{S}$); Otherwise we use $\mathcal{N}_{v_j}$ to represent the set of users who have interacted with item $v_j$. 


With the final node representations $\mathbf{e}_u$/$\mathbf{e}_v$, the goal of our primary social recommendation task is to predict how likely user $u_i$ would interact with item $v_j$ through an inner product: 
\begin{align}
    \hat{y}_{u_{i}v_{j}}= \mathbf{e}^T_{u_i}\mathbf{e}_{v_j}. 
\end{align}where $\hat{y}_{u_{i}v_{j}}$ denotes the predicted rating of user $u_i$ on item $v_j$ generated by the model.

\subsubsection{Self-supervised Auxiliary Tasks (SS-A Tasks)} 
\label{subsec:ssl_task}
In order to take advantage of rich unlabeled data in social recommendations, we propose to construct multiple self-supervised auxiliary (SS-A) tasks to provide self-supervision signals for assisting the training of primary recommendation task without using manual annotations.
According to the social correlation theories~\cite{milo2002network,newman2018networks,shi2016survey}, we explore multiple social SS-A tasks based on graph structure $\mathcal{G}=(\mathcal{G}^{\mathbf{R}}, \mathcal{G}^\mathbf{S})$. 

\begin{itemize}

\item \textbf{Triangle Relations} (Social Triangle, Joint Triangle): Triangle relations can reflect users preferences towards items and their social behaviors in social network~\cite{milo2002network,benson2016higher}, such as friendship, community of interest, as shown in Figure~\ref{high_order} (a) and (b). Here, we introduce two representative types of triangle relations, namely social triangle ($(\textbf{S}\textbf{S})\odot\textbf{S}$) and joint triangle ($(\textbf{R}\textbf{R}^T)\odot\textbf{S}$), where $\odot$ denotes the element-wise product. 

\item \textbf{$N$-hop Neighbors} (1, 2, 3, ..., n-hop neighbors): The $N$-hop neighbors auxiliary tasks explore the best way to propagate messages among users, where nodes in social networks can reach others via the shortest path between them~\cite{newman2018networks}. 

\item \textbf{Meta-paths}: Meta-paths~\cite{shi2016survey} have been widely used to capture diverse, long-range, and meaningful relations between entities in heterogeneous graph, where a path represents a sequence of nodes connected with heterogeneous edges. Here, in social recommendation, we  consider two types of meta paths to model social behaviors, i.e., $u_i \to v_j \to u_{i'}$ ($uvu$)  and $u_i \to u_{i'}  \to v_j \to u_{i''}$ ($uuvu$), where $u_i, u_{i'}, u_{i''} \in U$ and $v_j \in V$. More meta paths  will be explored in the future.
\end{itemize}

For simplicity, the set of SS-A tasks can be denoted as $\text{SSL}=\{\text{ssl}_1, \text{ssl}_2, ..., \\ \text{ssl}_s\}$, and the number of such tasks is $|\text{SSL}|$. 
The training data for SS-A tasks can be obtained from raw data without using any manual annotations. 
For instance, for $u_i \to v_j \to u_{i'}$ in meta-$uvu$, we need to distinguish  whether a pair of users (i.e., $(u_i, u_{i'})$) exists such relation belonging to $\text{ssl}_s$ (i.e., meta-$uvu$), where the self-supervised learning label $y_{u_i,u_{i'}}^{\text{ssl}_s} = 1$ if there exists such relation belonging to $\text{ssl}_s$, otherwise $y_{u_i,u_{i'}}^{\text{ssl}_s} = 0$. Then, the SS-A tasks' prediction can be formulated via inner product on users' final representations as follows:
\begin{align}
\hat{y}_{u_i,u_{i'}}^{\text{ssl}_{s}}=\mathbf{e}^T_{u_i}\mathbf{e}_{u_{i'}}.
\end{align}where $\hat{y}_{u_i,u_{i'}}^{\text{ssl}_{s}}$ denotes the model-predicted probability of a social relation between user $u_i$ and user $u_{i'}$.

\subsubsection{Optimization Loss for Primary Recommendation and SS-A Tasks}
With the supervised labels from the primary recommendation task and self-supervised labels from SS-A tasks, we employ the pairwise BPR loss~\cite{rendle2012bpr} to encourage the prediction of an observed entry to be higher than its unobserved counterparts:
\begin{small}
\begin{align}
    \mathcal{L}_{\text{pri}}= \sum_{(u_{i},v_{j},v_{j'}) \in O_{\text{pri}}}-\ln\sigma(\hat{y}_{u_{i}v_{j}}-\hat{y}_{u_{i}v_{j'}}) +\delta||\mathbf{E}^{(0)}||^2, \\
    \mathcal{L}_{\text{ssl}_s}= \sum_{(u_{i},u_{i'},u_{j}) \in O_{\text{ssl}_s}}-\ln\sigma(\hat{y}^{\text{ssl}_s}_{u_{i}u_{i'}}-\hat{y}^{\text{ssl}_s}_{u_{i}u_{j}}) +\delta||\mathbf{E}^{(0)}||^2, 
\end{align}
\end{small}where a hyper-parameter $\delta$ is used to control the contribution of $\mathcal{L}_2$ regularization $||\mathbf{E}^{(0)}||^2$ so as to reduce the risk of overfitting. $\sigma$ is the sigmoid function. As for primary recommendation task, the set of $O_{\text{pri}}=\{(u_{i},v_{j},v_{j'})|(u_{i},v_{j})\in \mathcal{E}_{\text{pri}},(u_{i},v_{j'}) \notin \mathcal{E}_{\text{pri}} \}$ denotes the pairwise training data in primary recommendation task, where $(u_{i},v_{j}) \in \mathcal{E}_{\text{pri}}$ indicates that the interaction between user $u_i$ and item $v_j$ is observed (positive) pair, while $(u_{i},v_{j'}) \notin \mathcal{E}_{\text{pri}}$ indicates that the interaction between user $u_{i}$ and item $v_{j'}$ is not observed (negative).
Similarly, as for BPR loss in SS-A task $\text{ssl}_s$ (i.e.,  $\mathcal{L}_{\text{ssl}_s}$),  the set of $O_{\text{ssl}_s}=\{(u_{i},u_{i'},u_{j})|(u_{i},u_{i'})\in \mathcal{E}_{\text{ssl}_s},(u_{i},u_{j}) \notin \mathcal{E}_{\text{ssl}_s} \}$ denotes the pairwise training data in SS-A task $\text{ssl}_s$, where $(u_{i},u_{i'}) \in \mathcal{E}_{\text{ssl}_s}$ indicates that the interaction between user $u_i$ and user $u_{i'}$ is observed (positive) pair, while $(u_{i},u_{j}) \notin \mathcal{E}_{\text{ssl}_s}$ indicates that the interaction between user $u_{i}$ and user $u_{j}$ is not observed (negative).
To optimize SS-A tasks, we apply negative sampling by randomly selecting an unobserved user/item and pairing it with the self-supervised social relations as a negative instance.

\subsection{Automatic Weighting Mechanism}
To effectively learn user and item representations, we present a joint optimization objective that combines the primary recommendation task with various SS-A tasks, which can be formally defined as:
\begin{align}
 \mathcal{L}_{\text{joint}}= \sum_{i=1}^{N}\frac{1}{N}\mathcal{L}^{\text{train}}_{\text{pri},i} +  \sum_{s=1}^{|\text{SSL}|}\lambda_{\text{ssl}_s}\sum_{i=1}^{N_{\text{ssl}_s}}\frac{1}{N_{\text{ssl}_s}} \mathcal{L}_{\text{ssl}_s,i}, \label{eq:ssl_ori_object}
\end{align}where $\mathcal{L}_{\text{pri},i}^{\text{train}}$ and $\mathcal{L}_{\text{ssl}_s,i}$ are the $i$-th instance of training set in primary recommendation task and SS-A task $\text{ssl}_s$, respectively. $N$ and $N_{\text{ssl}_s}$ denote the size of  training set in primary recommendation task and SS-A task $\text{ssl}_s$. $\lambda_{\text{ssl}_s}$ denotes the hyper-parameters to control the contribution on SS-A task $\text{ssl}_s$ to primary recommendation task, which can have a significant influence on model's performance in various recommendation scenarios and over different datasets. 
However, most existing works assign equal weight to $\lambda_{\text{ssl}_s}$ or heavily tune them in experiments, which can result in sub-optimal recommendation performance due to negative transfer among various SS-A tasks.

In our work, to address the above issue, AusRec aims to learn the primary recommendation task by automatically leveraging multiple SS-A tasks with flexible weights.
More specifically, an automatic weighting mechanism is introduced to automatically balance various SS-A tasks via learning (automatic) weights $\lambda_{\text{ssl}_s}$ for assisting representation learning in social recommendations.
This automatic weighting mechanism is formulated as a meta-learning task using bi-level optimization~\cite{wang2021bridging}, where the joint optimization objective of the primary social recommendation task with various SS-A tasks can be defined as: 
\begin{align}
    \min \limits_{\mathbf{W},\mathbf{\Theta}}\sum_{i=1}^{|D^{\text{pri-meta}}|}&\frac{1}{|D^{\text{pri-meta}}|}   \mathcal{L}^{\text{meta}}_{\text{pri},i} (\mathbf{W}^*(\mathbf{\Theta})), \\ 
    \text{s.t.} \textbf{W}^*(\mathbf{\Theta})  = &\mathop{\arg\min}_{\mathbf{W}}(\sum_{i=1}^{|D^{\text{pri}}|}\frac{1}{|D^{\text{pri}}|}\mathcal{L}_{\text{pri},i}^{\text{train+meta}}({\mathbf{W}}) +\nonumber \\
    \sum_{s=1}^{|\text{SSL}|}\sum_{i=1}^{N_{\text{ssl}_s}}&\frac{1}{N_{\text{ssl}_s}}  V([\mathcal{L}_{\text{ssl}_s,i};e_{\text{ssl}_s}];\mathbf{\Theta})\mathcal{L}_{\text{ssl}_s,i}({\mathbf{W}})), \label{eq:ssl_object}
\end{align}
\noindent where the automatic weights $\lambda_{\text{ssl}_s}$ can be learned by an automatic weighting network $V(\cdot;\mathbf{\Theta})$ which can be modeled as a two-layer MLP network with trainable parameters $\mathbf{\Theta}$.
$\mathbf{W}$ denotes the parameters of the recommendation encoder in both the primary task and the auxiliary tasks. 
$\mathcal{L}^{\text{meta}}_{\text{pri},i}$ is loss function of the $i$-th instance of meta data $D^{\text{pri-meta}}$ in the primary recommendation task. $D^{\text{pri-meta}}$ is a small portion of the primary task's training data $D^{\text{pri}} =D^{\text{train+meta}} = \{D^{\text{pri-train}}, D^{\text{pri-meta}}\}$ to guide the learning of the automatic weighting network. $\mathcal{L}^{\text{train+meta}}_{\text{pri},i}$ and $\mathcal{L}_{\text{ssl}_s,i}$ are loss functions of the $i$-th instance of training set in the primary recommendation task and SS-A tasks.
$D^{\text{SSL}}$ denotes the training set for SS-A task. $|D^{\text{pri}}|$ is the size of whole training set in the primary recommendation task, $|D^{\text{pri-meta}}|$ is the size of meta data from the primary recommendation task and much less than the size of the whole training set.

In each training epoch under meta learning paradigm~\cite{wang2021bridging,shu2019meta}, 
the automatic weighting network $V([\mathcal{L}_{\text{ssl}_s,i};e_{\text{ssl}_s}];\mathbf{\Theta})$
takes the concatenation of $\mathcal{L}_{\text{ssl}_s,i}$ and  $e_{\text{ssl}_s}$ as input, where $e_{\text{ssl}_s}$ is the one-hot vector of task $\text{ssl}_s$ to distinguish which task the loss of input to the automatic weighting network comes from. 
Note that the formulation in Eq.~\ref{eq:ssl_object} enables learning to support the primary task by optimizing $\mathbf{\Theta}$ through meta-learning.
The optimal parameters $\mathbf{\Theta}^*$ can be obtained by minimizing the following loss:
\begin{align}
    \mathbf{\Theta}^* &=\mathop{\arg\min}_{\mathbf{\Theta}}\mathcal{L}_{\text{pri}}^{\text{meta}}(\mathbf{W}^*(\mathbf{\Theta})) \\
    &\triangleq \sum_{i=1}^{|D^{\text{pri-meta}}|}{\frac{1}{|D^{\text{pri-meta}}|}}\mathcal{L}_{\text{pri},i}^{\text{meta}}(\mathbf{W}^*(\mathbf{\Theta})).
\end{align}

\subsection{Model Training}
The process to optimize our proposed AusRec is shown in the right part of  Figure~\ref{fig.2}, which includes three stages to update the parameters of recommendation encoder $\mathbf{W}$ and automatic weighting network $\mathbf{\Theta}$. Specially, stage I is a hypothetical update step for the main recommendation encoder based on the current task weights. Stage II uses the performance of this hypothetical encoder on a  meta dataset to evaluate how good the current weights are, and updating the weighting network accordingly. And stage III performs the actual update of the recommendation encoder using the newly improved weights. The overall training procedure of our proposed AusRec can be summarized in Algorithm~\ref{algor}.

\subsubsection{Stage I: Updating Parameters of Recommendation Encoder}

At each training epoch, we begin by sampling mini-batches of data from the primary recommendation task and SS-A tasks.
Then, the loss $\mathcal{L}_{\text{ssl}_s}$ of the SS-A tasks are weighted by the automatic weighting network and summed up with the loss $\mathcal{L}_{\text{pri}}^{\text{train}}$ of the primary recommendation task to obtain the final joint loss. 
To alleviate the difficulty of  the bi-level optimization ~\cite{wang2021bridging}, the parameters of recommendation encoder $\mathbf{W}^{(k)}$ can be approximated with the updated parameters $\hat{\mathbf{W}}^{(k)}$ in meta learning paradigm as follows:

\begin{small}
\begin{align}
\label{eql.15}
    \mathbf{\hat{W}}^{(k)}(\mathbf{\Theta})=\mathbf{W}^{(k)}-\alpha \bigtriangledown_{\mathbf{W}^{(k)}}&(\sum_{i=1}^{|D^{\text{pri-train}}|}\frac{1}{|D^{\text{pri-train}}|}\mathcal{L}_{\text{pri},i}^{\text{train}}(\mathbf{W}^{(k)})  \nonumber\\
   + \sum_{s=1}^{|\text{SSL}|}\sum_{i=1}^{N_{\text{ssl}_s}}\frac{1}{N_{\text{ssl}_s}}
    V([\mathcal{L}_{\text{ssl}_s,i}(\mathbf{W}^{(k)});&e_{\text{ssl}_s}];\mathbf{\Theta}^{(k)})\mathcal{L}_{\text{ssl}_s,i}(\mathbf{W}^{(k)})),
\end{align}
\end{small}where $\alpha$ is the learning rate for $\mathbf{W}$. 
$\mathcal{L}_{\text{pri},i}^{\text{train}}$ is the $i$-th instance loss function of data on $D^{\text{pri-train}}$.

\begin{algorithm}[t]
    \caption{The proposed AusRec Framework}\label{algor}
    \KwIn{Training data for the primary recommendation task $D^{\text{pri}}$ and SS-A tasks $D^{\text{SSL}}$, max iteration $K$, learning rate $\alpha$,$\beta$.}

    \KwOut{Social recommendation  parameters $\mathbf{W}^{K}$, and Automatic Weighting Network parameters $\mathbf{\Theta}^{K}$.}

    Initialization parameters $\mathbf{W}^1$,$\mathbf{\Theta}^1$; \\
    \For(){$k = 1 \dots K$}{
        Sample minibatch $D^{\text{pri}}_{n}$ from primary task training data $D^{\text{pri}}$;
        
        Sample minibatch $D^{\text{SSL}}_{n}$ from SS-A task training data $D^{\text{SSL}}$;
        
        Split $D^{\text{pri}}_{n}$ into $\{D^{\text{pri-train}}_{n}, D^{\text{pri-meta}}_{n} \}$; 
        
        Update $\mathbf{\hat{W}}^{(k)}$ by Eq. (\ref{eql.15}) with $D^{\text{pri-train}}_{n}$ and $D^{\text{SSL}}_{n}$;
        
        Update $\mathbf{\Theta}^{(k)}$ by Eq. (\ref{eql.16}) with $D^{\text{pri-meta}}_{n}$;
        
        Update $\mathbf{W}^{(k)}$ by Eq. (\ref{eql.17}) with $D^{\text{pri}}_{n}$ and $D^{\text{SSL}}_{n}$;
    }
\end{algorithm}

\subsubsection{Stage II: Updating Parameters of Automatic Weighting Network}


Once the parameters $\mathbf{W}^{(k)}$ of the recommendation encoder are updated, the parameters $\mathbf{\Theta}^{(k)}$ of the automatic weighting network can then be updated accordingly with the updated parameters $\mathbf{\hat{W}}^{(k)}$ on $D^{\text{pri-meta}}$, which is defined as follows: 
\begin{equation}
\label{eql.16}
    \mathbf{\Theta}^{(k+1)}=\mathbf{\Theta}^{(k)}-\beta \bigtriangledown_{\mathbf{\Theta}^{(k)}}\mathcal{L}_{\text{pri}}^{\text{meta}}(\mathbf{\hat{W}}^{(k)}(\mathbf{\Theta}^{(k)})),
\end{equation}where $\beta$ is the meta learning rate for updating $\mathbf{\Theta}$. We update parameters of the automatic weighting network on the meta data (i.e., $D^{\text{pri-meta}}$) from the primary recommendation task.  
This update allows softly select SS-A tasks and balances them for the primary recommendation task, so as to enhance the representation learning of users and items in social recommendations. 
Conversely, without such an automatic weighting network to balance various SS-A tasks on the joint loss (i.e., Eq.~\ref{eq:ssl_ori_object}), multiple SS-A tasks may result in negative knowledge transfer on the primary recommendation task via dominating the main training objective.

\subsubsection{Stage III: Meta Updating Parameters of Recommendation Encoder}

At last, with the updated parameters $\mathbf{\Theta}^{(k+1)}$ of the automatic weighting network,  the parameters $\mathbf{W}^{(k)}$ of recommendation encoder can be further updated through $D^{\text{pri}}=\{D^{\text{pri-train}}, D^{\text{pri-meta}}\}$ and  $D^{\text{SSL}}$ as follows:
\begin{small}
\begin{equation}
\label{eql.17}
    \begin{aligned}
    \mathbf{W}^{(k+1)}=\mathbf{W}^{(k)}-\alpha \bigtriangledown_{\mathbf{W}^{(k)}}(&\sum_{i=1}^{|D^{\text{pri}}|}\frac{1}{|D^{\text{pri}}|}\mathcal{L}_{\text{pri},i}^{\text{train+meta}}(\mathbf{W}^{(k)}) + \\
    \sum_{s=1}^{|\text{SSL}|}\sum_{i=1}^{N_{\text{ssl}_s}}\frac{1}{N_{\text{ssl}_s}}
    V(\mathbf{[\mathcal{L}}_{\text{ssl}_s,i}(\mathbf{W}^{(k)});&e_{\text{ssl}_s}];\mathbf{\Theta}^{(k+1)})\mathcal{L}_{\text{ssl}_s,i}(\mathbf{W}^{(k)})).
    \end{aligned}
\end{equation}

\vskip -0.2in
\end{small}

\subsection{Model Inferencing}
During inference, AusRec only retains the learned user and item representations and performs standard recommendation prediction. All self-supervised auxiliary tasks and the meta-learning optimization process are only activated during training and are completely removed at inference time. Therefore, the inference complexity and memory consumption are identical to those of conventional recommendation models.

\section{Experiment}
\label{sec:Experiments}
In this section, we conduct comprehensive experiments to validate the effectiveness of our proposed method.
We begin by presenting the experimental settings, followed by a comparison of our proposed method with various baselines on representative datasets. Finally, we perform detailed ablation studies on the proposed method.






\subsection{Experimental Settings}


\subsubsection{Datasets}
\label{sec:datasets}
To evaluate how effective the proposed method is, we carry out comprehensive tests using three representative datasets, including LastFM\footnote{\url{https://www.last.fm}}, Epinions\footnote{\url{http://www.epinions.com}}, and DBook\footnote{\url{https://book.douban.com}}. 
A statistical overview of these three datasets appears in Table~\ref{tab:dataset}.

\begin{itemize}
    \item \textbf{LastFM}. This dataset is provided by last.fm, a UK-based internet radio and music community. It contains data from social networks, making it a popular choice for testing social recommendation models. The dataset includes information on each user's favorite artists along with their play counts, as well as user-generated application tags that can be used to create content vectors. Specifically, the dataset comprises 1,880 users and 4,489 items.
    \item \textbf{Epinions}. This dataset is sourced from Epinions, a product review website where users can freely register and write subjective reviews on various items, including software, music, TV shows, hardware, and office appliances. The version of the dataset used in our proposed method consists of 5,957 users and 11,027 items. 
    \item  \textbf{DBook}. This dataset is collected from the Douban.Book website and includes users' book ratings (ranging from 1 to 5) as well as information about social networks between users. This is the largest of the three datasets used in this paper, containing 13,024 users and 22,347 items.
\end{itemize}



\begin{table}[!t]
\label{table3*}
\centering
\renewcommand\arraystretch{1}
\caption{Statistics of the experimental data.}
\label{tab:dataset}
\scalebox{0.9}
{
\begin{tabular}{c|cccccc}
    \toprule
    Dataset & users & items & Interactions& Density($\mathbf{R}$)& Connections& Density($\mathbf{S}$)\\
    \midrule
    LastFM& 1,880 & 4,489 & 52,668& 0.00624& 25,434& 0.00720\\ \midrule
    Epinions& 5,957 & 11,027 & 132,819& 0.00202& 98,959& 0.00277\\  \midrule
    DBook& 13,024 & 22,347 & 792,062& 0.01137& 169,150& 0.00100\\
    \bottomrule
\end{tabular} 
}
\end{table}

\subsubsection{Parameter Setting}
During training our proposed method, the parameters are set as follows: batch size is 2048, consistent with LightGCN. User and item embedding sizes $d$ are searched within the range $[128, 160, 192, 224, 256]$ with a step size of 32.
There are two learning rates in our proposed method, including the learning rate $lr$ for parameters updating of the target recommendation encoder, and the meta learning rate $mlr$ for parameters updating of the Automatic Weighting Network.
We select $lr$ from $[0.001, 0.005, 0.01, 0.05]$ and $mlr$ from $[0.0001, 0.0005, 0.001, 0.005]$, using a grid search to identify the optimal parameters. The optimal configuration is $lr = 0.001$ and $mlr = 0.0001$. We divide the interactions into training and testing sets at a $4:1$ ratio.
Specific experiment is described below. 
Besides, we set the ratio of meta data in  dataset as $5\%$. 
Adam optimizer is employed for updating the model's parameters. To ensure a fair comparison, we adhere to the default settings of the baselines. All experiments are performed on a system with an Intel(R) Xeon(R) Gold 5120 CPU and an NVIDIA Titan RTX 2080Ti 24G GPU.

\begin{table*}[tp]
\renewcommand{\arraystretch}{1.3}
\centering
\caption{Performance comparison of different recommendation methods.}
\label{tab:comparsion_all}
\scalebox{0.46}
{
\begin{tabular}{c|l|ccccccccccc|c}
\toprule
\multicolumn{1}{c|}{\textbf{Datasets}}              & \multicolumn{1}{c|}{\textbf{Metrics}} & 
\multicolumn{1}{c}{\textbf{MF}} & \multicolumn{1}{c}{\textbf{NeuMF}}& \multicolumn{1}{c}{\textbf{NGCF}}& \multicolumn{1}{c}{\textbf{LightGCN}} & \multicolumn{1}{c}{\textbf{LightGCN-S}}& \multicolumn{1}{c}{\textbf{SocialMF}} & \multicolumn{1}{c}{\textbf{DiffNet}} & \multicolumn{1}{c}{\textbf{SGL}}& \multicolumn{1}{c}{\textbf{MHCN}} & \multicolumn{1}{c}{\textbf{HGCL}} & \multicolumn{1}{c}{\textbf{AusRec}}&\multicolumn{1}{|c}{\textbf{Improv.}}\\ \midrule
\multirow{6}{*}{\textbf{LastFM}} & \textbf{Recall@5}       &    0.0786  &0.0835      &      0.0950&  0.1225    &     0.1254    &  0.1003    &     0.0908      &0.1261  &     0.1237    &  \underline{0.1268}   & \textbf{0.1313} &  3.55\%     \\  
& \textbf{Recall@10}    &      0.1290    & 0.1281    &     0.1483 &    0.1925   &  0.1925     &   0.1538     &    0.1425        &0.1920&       0.1944      & \underline{0.1964} &    \textbf{0.2029} & 3.31\%     \\  
& \textbf{Recall@20} &       0.1943  &0.1873    &     0.2215 &  0.2748   &    0.2774    &   0.2360    &    0.2144        &  \underline{0.2818}&    0.2789     &  0.2816  &    \textbf{0.2909} & 3.23\%         \\
\cline{2-14}
                                 & \textbf{NDCG@5}    &   0.1034    &0.1183       & 0.1223  &  0.1640       &   0.1678    & 0.1313        &   0.1297         &     0.1687& 0.1653           & \underline{0.1690}  & \textbf{0.1734} & 2.60\%     \\  
                                 & \textbf{NDCG@10}     &     0.1156    &0.1316     &     0.1371 &  0.1791   &   0.1814     &  0.1423    &     0.1442      &      0.1808& 0.1818        &    \underline{0.1833}&\textbf{0.1889} & 3.06\%         \\ 
                                 & \textbf{NDCG@20}      &    0.1426    &0.1601      &    0.1670&   0.2148    &   0.2152   &   0.1771     &     0.1781      &0.2189& 0.2191        &    \underline{0.2200}&\textbf{0.2244} & 2.00\%         \\ \midrule
\multirow{6}{*}{\textbf{Epinions}} & \textbf{Recall@5}       &    0.0191   &  0.0192  &      0.0309&  0.0372    &     0.0364   &   0.0278    &     0.0259      & 0.0341&      0.0313         & \underline{0.0376}&\textbf{0.0421} & 11.99\%     \\  
                                & \textbf{Recall@10}    &   0.0306   &   0.0339     &    0.0501&   0.0585     &   0.0574    &   0.0450     &    0.0416        &    0.0531& 0.0523          &   \underline{0.0586}&\textbf{0.0645}&  10.07\%     \\  
                                & \textbf{Recall@20} &    0.0470   &  0.0567   &    0.0777 &  0.0908    &    0.0904   &     0.0679  &    0.0669        &    0.0796& 0.0823          & \underline{0.0913}  &  \textbf{0.1011}&  10.73\%       \\ \cline{2-14}
                                 & \textbf{NDCG@5}    &     0.0164    &  0.0185   &     0.0251&  0.0319     &  0.0321      &   0.0237     &    0.0261        &   0.0304&    0.0338         &   \underline{0.0346}&\textbf{0.0365} &  5.49\%     \\  
                                 & \textbf{NDCG@10}     &     0.0208    &  0.0250   &     0.0324&  0.0398    &   0.0399    &    0.0302    &     0.0330      &  0.0373&     0.0430        &   \underline{0.0441} &\textbf{0.0447}  
 & 1.36\%         \\
                                 & \textbf{NDCG@20}      &    0.0259   &   0.0336    &   0.0421&   0.0505     &  0.0508     &  0.0378     &    0.0413       &   0.0461&       0.0541   &     \underline{0.0552} &\textbf{0.0564} &    2.17\%       \\ \midrule
\multirow{6}{*}{\textbf{DBook}} & \textbf{Recall@5}       &    0.0621   &  0.0428  &      0.0649 &  0.0734  &     \underline{0.0762}   &   0.0570    &     0.0556      &  0.0729 &     0.0749     &  0.0758  & \textbf{0.0826}  &8.40\%     \\ 
                                & \textbf{Recall@10}    &   0.0922     &  0.0723    &    0.0976 &   0.1105    &   0.1172     &   0.0864    &    0.0870        & 0.1043  &  0.1122          &   \underline{0.1174}&\textbf{0.1218}  & 3.75\%     \\ 
                                & \textbf{Recall@20} &    0.1355    &  0.1154  &    0.1427 &   0.1637   &    \underline{0.1716}    &  0.1336    &    0.1290        &0.1431  &   0.1670          & 0.1712   & \textbf{0.1721}   &0.29\%       \\ \cline{2-14}
                                 & \textbf{NDCG@5}    &     0.0596   &  0.0421    &     0.0625&    0.0752   &  0.0779     &     0.0578    &    0.0602        & 0.0513   &   0.0771         &  \underline{0.0813} &\textbf{0.0865} &  6.40\%     \\  
                                 & \textbf{NDCG@10}     &     0.0677    &   0.0532  &     0.0727 & 0.0854    &   0.0900     &   0.0666   &     0.0728      &  0.0618  &   0.0872        & \underline{0.0904}  & \textbf{0.0968} &   7.08\%         \\  
                                 & \textbf{NDCG@20}      &    0.0813    &  0.0690    &   0.0870 &  0.1024     &  0.1069    &   0.0815     &    0.0876       &   0.0723    &   0.1058      & \underline{0.1100} &  \textbf{0.1135} & 3.18\%       \\ 
                \bottomrule

\end{tabular}
}
\end{table*}

\subsubsection{Baselines}
To assess effectiveness, we compare the proposed method against three categories of representative baselines, covering recommendation models without social network information (\textbf{MF}~\cite{koren2009matrix}, \textbf{NeuMF}~\cite{he2017neural}, \textbf{NGCF}~\cite{wang2019neural}, \textbf{LightGCN}~\cite{he2020lightgcn}), social  recommendation systems (\textbf{SocialMF}~\cite{jamali2010matrix}, \textbf{LightGCN-S}, \textbf{DiffNet}~\cite{wu2019neural}), and self-supervised recommendation systems (\textbf{SGL}~\cite{wu2021self}, \textbf{MHCN}~\cite{yu2021self}, \textbf{HGCL}~\cite{chen2023heterogeneous}). These baseline methods are summarized as follows:
\begin{itemize}
\item MF~\cite{koren2009matrix}: This method stands as a classic collaborative filtering approach, utilizing only user-item direct interactions as the primary input.
\item NeuMF~\cite{he2017neural}: This approach represents a cutting-edge matrix factorization model built on a neural network architecture.

\item NGCF~\cite{wang2019neural}: This is the first GNN-based model to integrate high-order connectivity between users and items.
\item LightGCN~\cite{he2020lightgcn}: This is a collaborative filtering GNNs-based recommendation method  by simplifying NGCF model~\cite{wang2019neural}, achieving promising recommendation performance. 

\item LightGCN-S: This model improves upon LightGCN by integrating social
network information.



\item SocialMF~\cite{jamali2010matrix}: This method integrates trust information propagation into the MF model for recommendation systems.

\item DiffNet~\cite{wu2019neural}: This GCN-driven social recommendation approach captures recursive and dynamic social influence across both user and item domains.
\item SGL~\cite{wu2021self}: This advanced self-supervised recommendation method employs diverse data augmentation techniques, such as node drop, edge drop, and random walk, to enhance recommendation performance.
\item MHCN~\cite{yu2021self}: This recent GCN-based social recommendation method incorporates self-supervised learning into hypergraph convolutional networks.
\item HGCL~\cite{chen2023heterogeneous}: This is a contrastive learning model dynamically applies knowledge from side information to model user-item interactions.
\end{itemize}

\subsection{Performance Comparison of Recommender Systems}
We begin by comparing the performance of our proposed method, AusRec, with all baseline methods. Table~\ref{tab:comparsion_all} presents the overall performance comparison based on Recall@5/10/20 and NDCG@5/10/20 metrics across three datasets. From these results, we observe the following:

\begin{itemize}

\item Our proposed AusRec demonstrates strong performance across three datasets, outperforming baseline methods on both metrics.
For example, our proposed method improves one of the most effective baselines LightGCN at least 10\% under Recall metric on Epinions dataset, and also outperforms social recommendation methods (i.e., LightGCN-S and SocialMF) and SSL methods (i.e., SGL and MHCN) on three datasets. 
This improvement highlights the effectiveness of our proposed method and the adaptability of the automatic weighting mechanism to various SS-A tasks. 


\item NGCF and LightGCN as representative GNNs-based recommendations can show much better performance than traditional recommendation methods (i.e., MF). On the other hand, compared with social recommendations, their performance is still competitive, indicating the power of GNNs in representation learning for recommendations.  

\item SGL as a general self-supervised recommendation method can achieve promising performance in some cases, though no social information is enhanced. This can be attributed to the integration of self-supervision signals. This observation demonstrates the effectiveness of self-supervised learning techniques in supporting the primary recommendation task.


\item Although MHCH and HGCL are state-of-the-art self-supervised social recommendation methods, they do not achieve the best performance across all cases.
Our proposed method AusRec can outperform all the baseline, which is attributed to considering the automatic weights on various SS-A tasks.

\item AusRec achieves varying degrees of improvement over baseline methods across the three datasets. Specifically, the performance gain is the smallest on the LastFM dataset, with a 3.55\% improvement in Recall@5, while the largest improvement is observed on the Epinions dataset, reaching 11.99\% in Recall@5. This variation may be attributed to the differing densities of the underlying social networks. Excessively high network density (LastFM) may introduce considerable noise in social relations, thereby degrading the quality of auxiliary tasks and consequently affecting the performance of the primary recommendation task. Conversely, overly sparse social networks (DBook) may fail to adequately capture users’ preferences, which can likewise hinder the overall effectiveness of the model.

\end{itemize}

In summary, our proposed method AusRec achieves significant performance improvements over various baselines, demonstrating the effectiveness of the proposed framework. The superior performance of AusRec can be attributed to its ability to incorporate multiple types of social relations as additional knowledge, which enriches the representation learning process. Moreover, the automatic weighting network effectively balances different self-supervised auxiliary tasks with the primary recommendation objective, enabling the model to capture more comprehensive user–item interactions and further enhance overall recommendation accuracy.

\subsection{Ablation Study}
In this part, to gain a deeper understanding of our framework, AusRec, we examine the impact of its components and hyper-parameters through comprehensive experiments.

 \subsubsection{The Change of Automatic Weights on Various SS-A Tasks during Training}
 \label{sec:5.3.1}
To grasp the workings of our proposed AusRec, we now analyze the change of automatic weights $\lambda_{\text{ssl}_s}$ on various SS-A tasks during the model training process. Note that weights $\lambda_{\text{ssl}_s}$ (learned by the automatic weighting network $V(\cdot;\mathbf{\Theta})$) aim to control the contribution of SS-A tasks $\text{ssl}_s$ on primary recommendation task, which can affect model's performance significantly.

\begin{figure*}[t]
  \centering
  \includegraphics[width=0.990\linewidth]{Weight.png}
  \caption{The change of automatic weights on various SS-A tasks during training (Best viewed in color).}
\label{weight trend}
\end{figure*} 

Figure~\ref{weight trend} presents the experimental results, which reveal that (1) the automatic weights on most SS-A tasks are unstable in the early training stage, while they can converge to a specific value at the end; and  (2) the weights on different SS-A tasks are different over three datasets. 
More specifically, the weight on SS-A task $\text{ssl}_2$ (i.e., joint triangle relation) is much largest than  that of other SS-A tasks  on Epinions dataset, while the weight on SS-A task $\text{ssl}_7$ (i.e., meta-path $uuvu$) contributes more on primary recommendation task  than that of other SS-A tasks on LastFM dataset.  
These observations suggest that the proposed automatic weighting mechanism can automatically control the contribution of various SS-A tasks across different datasets, so as to transfer self-supervision signals in various SS-A tasks to enhance social recommendation.

\begin{table}[t]
\renewcommand{\arraystretch}{1.3}
\centering
\caption{Results of AusRec with the same weight towards SS-A tasks.}
\label{tab:ablation experiment}
\scalebox{0.54}
{
\begin{tabular}{c|ccc|ccc|ccc}
\toprule

\textbf{Dataset}&\multicolumn{3}{c|}{\textbf{LastFM}}&\multicolumn{3}{c|}{\textbf{Epinions}} &\multicolumn{3}{c}{\textbf{DBook}}\\
\cmidrule{1-10}
\multicolumn{1}{c|}{\textbf{Metrics}}  & \multicolumn{1}{c}{\textbf{Recall@5}}& \multicolumn{1}{c}{\textbf{Recall@10}}& \multicolumn{1}{c}{\textbf{NDCG@10}}&  \multicolumn{1}{|c}{\textbf{Recall@5}}& \multicolumn{1}{c}{\textbf{Recall@10}}& \multicolumn{1}{c}{\textbf{NDCG@10}}& \multicolumn{1}{|c}{\textbf{Recall@5}}& \multicolumn{1}{c}{\textbf{Recall@10}}& \multicolumn{1}{c}{\textbf{NDCG@10}}\\
\midrule
\multirow{1}{*}{\textbf{AusRec}}  &  \textbf{0.1313}
 & \textbf{0.2029} & \textbf{0.1889} & \textbf{0.0421} & \textbf{0.0645}& \textbf{0.0447} & \textbf{0.0826} & \textbf{0.1218} & \textbf{0.0968}\\
\multirow{1}{*}{\textbf{w/o AW}}& 0.1213  & 0.1873 & 0.1758 & 0.0364 & 0.0589 & 
 0.0409 & 0.0705 & 0.1050 & 0.0841\\ \midrule
                \multirow{1}{*}{\textbf{Improv.}}
& 8.24\%  & 8.33\% & 7.45\% & 15.66\% & 9.51\% & 9.29\% & 17.16\% & 16.00\% & 15.10\%\\
\bottomrule
\end{tabular}
}
\end{table}

\subsubsection{Effect of Automatic Weighting Mechanism}
\label{sec:5.3.2}
To further study the effect of the automatic weighting mechanism, the weights of seven kinds of  SS-A tasks are equally set to 1 without differentiating the importance of SS-A tasks (w/o AW for short).  
Table~\ref{tab:ablation experiment} provides the experimental results across three datasets, showing that the model without the automatic weighting network (i.e., w/o AW) experiences a significant drop in recommendation performance compared to the proposed AusRec.
Therefore, we can see that removing the automatic weighting network can degrade performance significantly on three datasets,  indicating that the automatic weighting mechanism can contribute to automatically leveraging various auxiliary tasks for enhancing social recommendation.

\begin{figure*}[t]
\centering
{\subfigure[Recall@10]
{\includegraphics[width=0.49\linewidth]{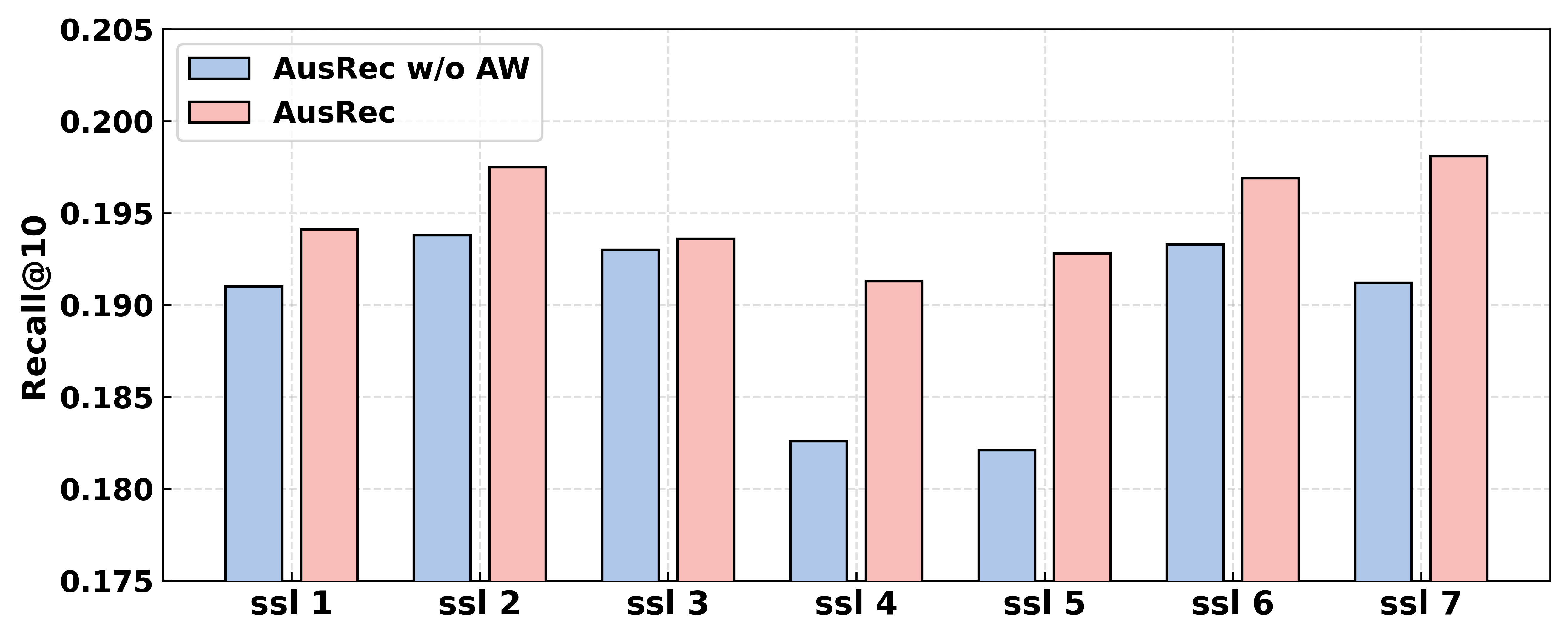}}}
{\subfigure[NDCG@10]
{\includegraphics[width=0.49\linewidth]{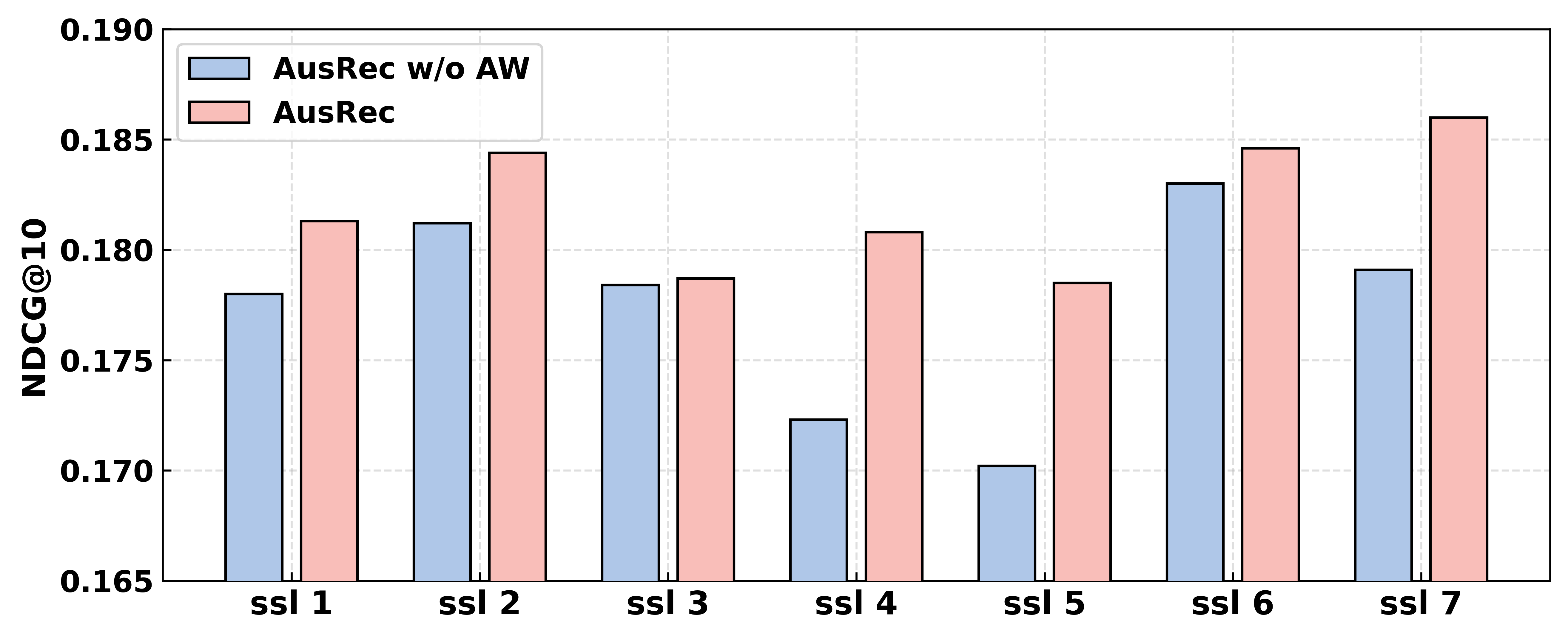}}}

\caption{Results of ablation experiments (single auxiliary task).}
\label{fig:one_aux}
\end{figure*}

\subsubsection{Effect of automatic weighting mechanism for only One SS-A task}
\label{sec:5.3.3}
To further study the effect of the automatic weighting mechanism, we use the automatic weighting mechanism with only one SS-A task, while the weight on variant (w/o AW)  towards each SS-A task is set to 1 without differentiating the important weights of the samples in the SS-A task. 
Here, we run the experiment on the LastFM dataset, as similar patterns are observed in other datasets. Figure~\ref{fig:one_aux} shows that the automatic weighting network positively impacts model performance with only one SS-A task under the evaluation metrics Recall@10 and NDCG@10.
For instance, among the seven SS-A tasks, the model with the automatic weighting network on SS-A task $\text{ssl}_5$ can achieve the largest performance improvement compared to the model without the automatic weighting network. 
And the model with SS-A task $\text{ssl}_3$ has only a slight performance improvement compared to that of the model without the automatic weighting network. 
Note that $\text{ssl}_3$ and $\text{ssl}_5$ denote the SS-A tasks based on 1-hop and 3-hop social relations, respectively. 
The reason for these observations is that SS-A task $\text{ssl}_5$ can explore the high-order signals (3-hop) in  social networks, hence providing more performance improvement by learning better representations, while SS-A task $\text{ssl}_3$ only explores the shallow signals (1-hop) in  social networks, which provides  limited signals to learn social information for enhancing the performance of social recommendation.
Besides, the observations mentioned above also suggest the model is not just learning a single weight per task, but is learning to weight per-instance losses dynamically, which is a more sophisticated form of optimization.
In summary, the phenomenon indicates that the automatic weighting network not only automatically selects SS-A tasks to improve the primary recommendation task, but also improves the impact of a single SS-A task on the primary recommendation task. 

\begin{figure*}[t]
\centering
{\subfigure[]
{\includegraphics[width=0.49\linewidth]{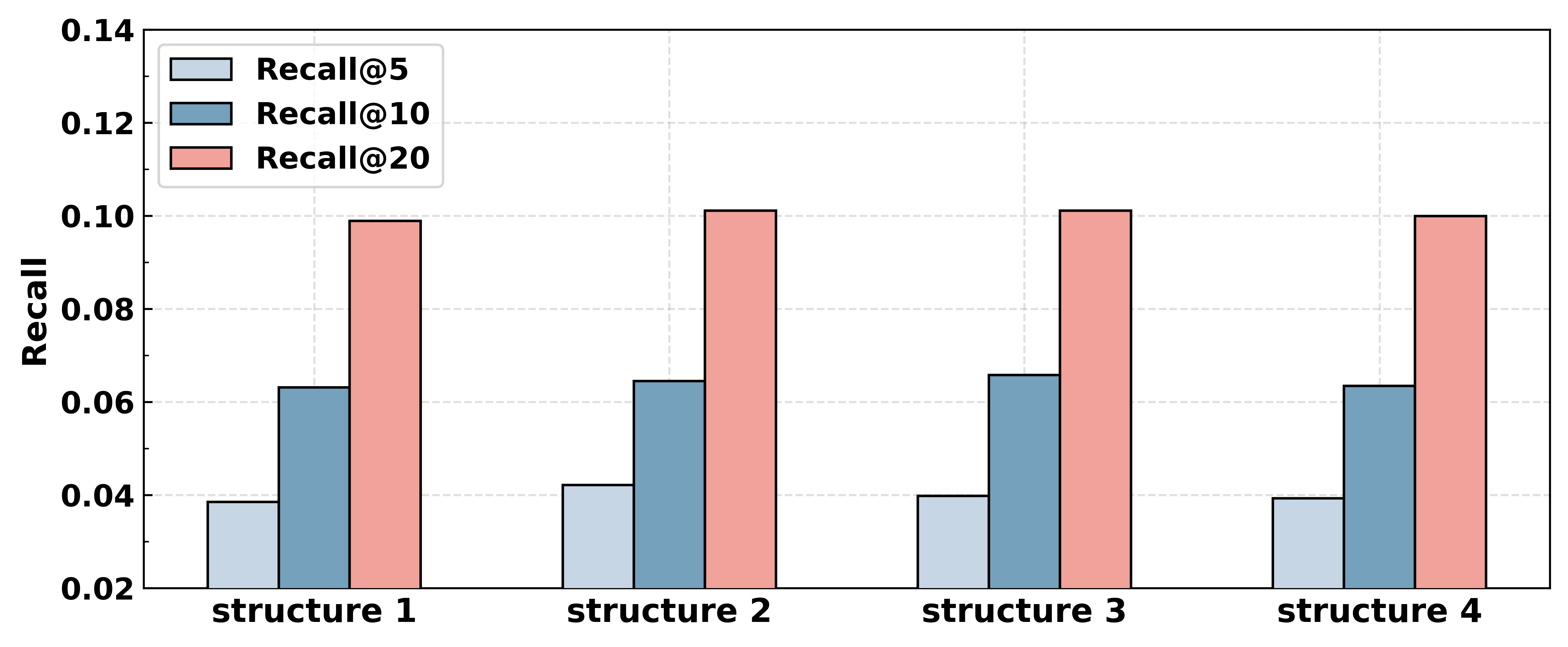}}}
{\subfigure[]
{\includegraphics[width=0.49\linewidth]{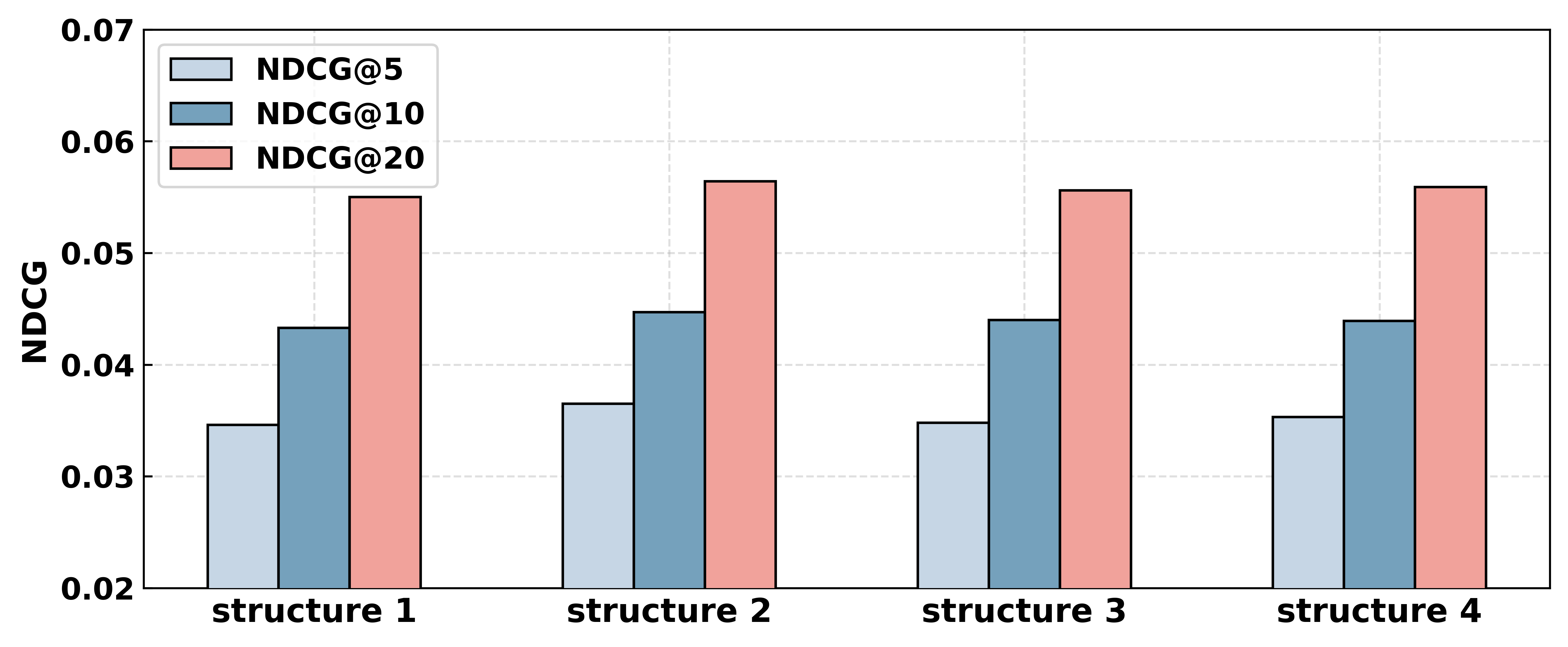}}}

\caption{Effect of Automatic Weighting Network Architecture.}
\label{fig:ablation experiment for different architcture}
\end{figure*}

\subsubsection{Effect of Automatic Weighting Network Architecture} 
\label{sec:5.3.4}
To examine the impact of the MLP architecture in the automatic weighting network, we conduct experiments on Epinions dataset with various MLP architecture configurations. Structures 1, 2, 3, and 4 correspond to the four structure configurations: 8-500-1, 8-1000-1, 8-1000-1000-1, and 8-100-100-1, respectively.
Figure~\ref{fig:ablation experiment for different architcture} presents the experimental results, showing that the network structure configuration 8-1000-1 achieves the best performance among the four configurations Epinions datasets.
For example, the network structure setting 8-1000-1 achieves the best performance under NDCG@10, while the rest of the three network structure settings only achieve suboptimal performance. Therefore, we select the network structure setting 8-1000-1 as the network structure setting for the automatic weighting network in our proposed method, achieving higher recommendation performance.

\begin{table*}[h]
\centering
\renewcommand\arraystretch{1}
\caption{Average training time per epoch (sec/epoch) on the LastFM dataset.}
\vskip -0.1in
\label{tab:time_comparison}
\scalebox{0.87}{
\begin{tabular}{c|ccccccc}
    \toprule
    Model & LightGCN & DiffNet &SGL&MHCN&HGCL&AusRec&AusRec(one)\\
    \midrule
    sec/epoch& 0.73 & 1.02 & 1.27& 2.43 &3.87&4.1&2.1\\
    \bottomrule
\end{tabular}}
\end{table*}

\subsection{Training Efficiency}
To evaluate the training efficiency of our model, we compared the time required to train our model for one epoch with that of other baseline models.
The experimental results are presented in Table~\ref{tab:time_comparison}. As shown in the table, LightGCN and DiffNet require the shortest training time per epoch, since they are trained solely on graph-structured recommendation data without incorporating any SS-A tasks.
SGL and MHCN incorporate different numbers of SS-A tasks, which consequently lead to an increase in their training time.
HGCL introduces multiple auxiliary tasks and employs a more complex network architecture, thereby further increasing the training time.
Our proposed model AusRec incorporates the largest number (seven) of auxiliary tasks among all compared approaches and adopts a meta-learning training paradigm. Consequently, its training efficiency is roughly comparable to that of HGCL. It is noteworthy that the training efficiency of AusRec can be further improved by reducing the number of SS-A tasks. As illustrated in Table~\ref{tab:time_comparison}, AusRec (one) represents the variant trained with only a single auxiliary task, requires approximately half the training time per epoch compared with the full version of the model.
Improving the training efficiency of the model is also an important direction for our future research.
\section{Limitation and Future Work}
Although AusRec achieves strong performance across various datasets, it still has several limitations that deserve further investigation.
First, the auxiliary tasks introduced in this work are relatively limited in diversity, as they are primarily constructed from one-hop or multi-hop user–user relations. Such relationships may not be sufficient to capture the full spectrum of user intentions and social influences in real-world scenarios. Second, the framework involves multiple auxiliary tasks and relies on meta-optimization for automatic weighting, which inevitably increases computational complexity compared with single-task recommendation models.

In future work, we plan to explore more diverse and semantically rich social relations (e.g., group-level, contextual, or temporal social interactions) to construct auxiliary tasks that better reflect users’ multifaceted preferences. We also aim to design more efficient optimization strategies to mitigate computational costs while preserving or enhancing recommendation accuracy.

\section{Conclusion}
\label{sec:conclusion}

In this work, we proposed an automatic self-supervised learning framework for social recommendations (\textbf{AusRec}), which can automatically balance the importance of various Self-Supervised Auxiliary (SS-A) tasks to enhance representation learning in social recommendations. 
What's more, our model provides advanced components with two following advantages: 
(1) Our proposed method can incorporate many SS-A tasks for social recommendations, which can take advantage of different kinds of social relation information across different datasets and downstream tasks; 
(2) The proposed automatic weighting mechanism facilitates the soft selection and balancing of SS-A tasks with the primary recommendation task, thereby enhancing user and item representation learning in social recommendations.
Extensive experiments on three real-world datasets demonstrate the effectiveness of our proposed method. In addition, the ablation studies also demonstrate that the automatic weighting mechanism can adapt multiple SS-A tasks for improving the primary recommendation task. 

\section{Acknowledgments}
This work was supported by a grant from the National Natural Science Foundation of China under grants (No.62372211), and the Science and Technology Development Program of Jilin Province (No.20250102216JC).




 \bibliographystyle{elsarticle-num} 
 \bibliography{main}





\end{document}